\documentclass[prb,aps,floats,floatsfix,twocolumn,superscriptaddress,showpacs]{revtex4}
\usepackage{amsfonts}
\usepackage{amsmath}
\usepackage{amssymb}
\usepackage{color,graphicx}
\usepackage{float}

\begin{document}
\title{Polarization dependence of semiconductor exciton and biexciton contributions to
phase-resolved optical two-dimensional Fourier-transform spectra}
\author{Alan~D.~Bristow}
\address{JILA, University of Colorado \& National Institute of Standards and
Technology, Boulder CO 80309-0440}
\author{Denis~Karaiskaj}
\address{JILA, University of Colorado \& National Institute of Standards and
Technology, Boulder CO 80309-0440}
\author{Xingcan~Dai}
\address{JILA, University of Colorado \& National Institute of Standards and
Technology, Boulder CO 80309-0440}
\author{Richard.~P.~Mirin}
\address{National Institute of Standards and
Technology, Boulder CO 80305}
\author{Steven~T.~Cundiff}
\email{cundiffs@jila.colorado.edu}
\address{JILA, University of Colorado \& National Institute of Standards and
Technology, Boulder CO 80309-0440}

\begin{abstract}
We study the coherent light-matter interactions associated with
excitons, biexcitons and many-body effects in GaAs quantum wells.
For most polarization configurations the phase-resolved
two-dimensional Fourier-transform (2DFT) spectra are dominated by
excitonic features, where their strength and dispersive lineshapes
is due to many-body interactions. Cross-linear excitation
suppresses many-body interactions, changing the lineshape and
strength of the 2DFT features.
\end{abstract}

\date{\today}
\maketitle

The coherent response of excitons in semiconductor quantum wells
(QWs) is strongly dependent on the excitation conditions and
material properties, such as polarization configuration and
inhomogeneous broadening (due to well-width fluctuations).
Contributions to the light-matter interactions include the
excitons themselves, the formation of excitonic
``\emph{molecules}," or biexcitons, and the many-body interactions
of these states. (See for example, the recent reviews
\onlinecite{CundiffOpEx,Axt1} and references therein.) The
interplay of these contributions has been explored though
intensity- and polarization-dependent transient four-wave mixing
(TFWM)
studies.\cite{Cundiff1,Yaffe,Bennhardt,Saiki,Kim,Schmitt-Rink,Patkar,Robart,Bott,Wang,Hu,Smirl,Shacklette,Mayer,Schafer,Adachi,Nickolaus,Langbein}
The latter result in changes of the dephasing
time,\cite{Cundiff1,Yaffe,Bennhardt,Saiki} the temporal profile of
the emission,\cite{Cundiff1,Bennhardt,Kim} and a phase shift of
the beats.\cite{Bennhardt,Schmitt-Rink} Some experiments have also
characterized the Stokes parameters of the emission with detailed
polarimetry.\cite{Patkar,Robart} Explanations of these results
vary and include inhomogeneity\cite{Cundiff1,Bennhardt} or
exciton-exciton interactions,\cite{Kim,Bott} such as
exciton-exciton exchange,\cite{Bennhardt,Robart}
excitation-induced dephasing (EID),\cite{Patkar,Wang,Hu,Smirl}
local-field corrections,\cite{Patkar,Hu} and excitation-induced
shift (EIS).\cite{Shacklette} Many authors have attributed the
polarization dependence to biexcitons and their subsequent
interactions.\cite{Yaffe,Saiki,Mayer,Schafer,Adachi,Nickolaus,Langbein}

TFWM measurements have not resulted in a completely satisfactory
explanation of the polarization-dependent coherent response,
because of ambiguities associated with competing processes in the
coherent response. Additional information has been gained by
recording the time evolution of the
emission.\cite{Cundiff1,Schneider} However, great enhancements are
obtained by explicitly tracking the evolving phase of the TFWM
signal using either a coherent-control scheme\cite{Breunig,Voss}
or two-dimensional Fourier-transform (2DFT)
spectroscopy.\cite{Li,Langbein2,Zhang2} The latter results in a
two-dimensional spectrum from the Fourier-transform of the phase
evolution of the signal along two time dimensions, and has
separated the population from coupling
contributions,\cite{Li,Langbein2} confirmed EID and EIS,\cite{Li}
and shown that agreement with theory requires the inclusion of
terms beyond the Hartree-Fock approximation.\cite{Zhang2}

In this paper, 2DFT spectroscopy is used to separate and isolate
the competing intraactions and interactions of the excitons and
biexcitons, which are strongly polarization dependent. Through a
quantitative comparison of the magnitude of 2DFT data and the
lineshape in the phase-resolved spectra, the selection rules are
exploited to demonstrate the suppression of either many-body or
biexcitonic effects in the coherent response. Clear indication of
the associated contributions are observed in the 2DFT spectra,
whereas they had only previously been inferred in TFWM
experiments. Many-body interactions are observed for most
excitation conditions as strong population and coherent coupling
peaks, and as dispersive lineshapes in the real part of the 2DFT
spectra. When many-body interactions are suppressed however, the
exciton and biexciton contributions are similar in strength and
the off-diagonal coupling peaks nearly vanish.

\begin{figure}[b]
\centering{\includegraphics[width=8.5cm]{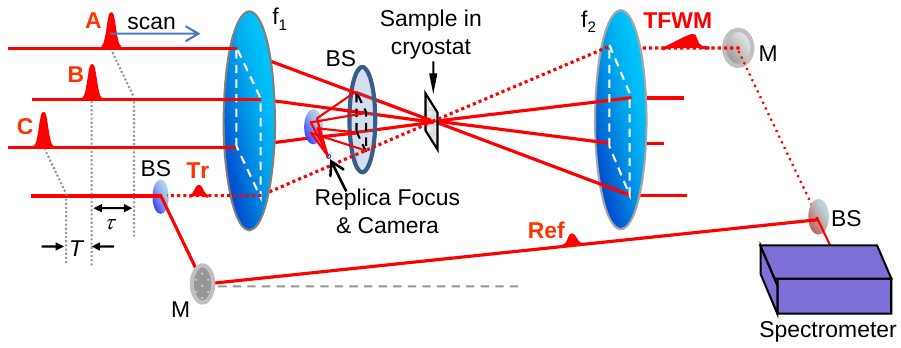}}
\caption{(Color online) Schematic diagram of the experimental
setup for optical two-dimensional Fourier-transform spectroscopy.
Notation: $\mathrm{f_i}$: lens; BS: beam spitter; M: mirror.}
\label{fig:fig1}
\end{figure}


A schematic diagram of the experimental setup is shown in
Fig.~\ref{fig:fig1}. Experiments are performed in the box geometry
with a mode-locked Ti:sapphire laser as the source. Pulses are
$\sim $200~fs and centered around 800~nm. Pulses are split into
four identical copies within a stabilized set of cascaded,
phase-stabilized and folded interferometers. The TFWM signal is
heterodyne detected with a phase-stabilized reference pulse that
is routed around the sample. The signal and reference are
collinearly recombined and recorded using spectral interferometry.
In this geometry there are three time periods, $\tau $ between the
first and second pulse, $T$ between the second and third pulse and
$t$ between the third pulse and the TFWM emission. S$_{I}$($\omega
_{\tau}$,$T$,$\omega _{t}$) 2DFT spectra are the Fourier transform
projections of the first and third time periods $\tau$ and $t$,
measured with the phase-matching condition
$k_{s}=-k_{A}+k_{B}+k_{C}$. Both the TFWM and 2DFT data are
acquired by scanning pulse A as indicated in Fig.~\ref{fig:fig1}.
These spectra are known as ``rephasing'' because dephasing due
inhomogeneous broadening is cancelled, which results in a photon
echo. Phase-matching conditions $k_{B}-k_{A}+k_{C}$ and
$k_{B}+k_{C}-k_{A}$ correspond to different time ordering of the
pulses. These techniques isolate the one- and two-photon
``non-rephasing" (non-echo like) coherent contributions, which are
not discussed in this paper.

2DFT experiments require sub-cycle phase-tracking and
stabilization. However, experimentally introduced phase shifts mix
the real and imaginary parts of the complex spectrum in the
as-measured data. The ``global phase," associated with the
nonlinear polarization in the sample, has previously been
determined by comparison of the phase-resolved TFWM to the
spectrally resolved transient absorption (SRTA).\cite{Jonas} For
cross-polarized excitation however, no corresponding configuration
of the SRTA exists. In this experiment the global phase is
determined by an \emph{in situ} all-optical method,\cite{Bristow2}
which involves: a) measuring the relative phases of the pump
($A$,$B$,$C$) and tracer (Tr) pulses through spatial interference
patterns at a replica focus shown in Fig.~\ref{fig:fig1}; b)
capturing the spectral phase of the TFWM signal, measured with the
heterodyne reference pulse (ref); and c) acquisition of the
spectral phase of the Tr pulse, also measured with the Ref pulse
by spectral interferometry.

\begin{figure}[b]
\centering{\includegraphics[width=7.0cm]{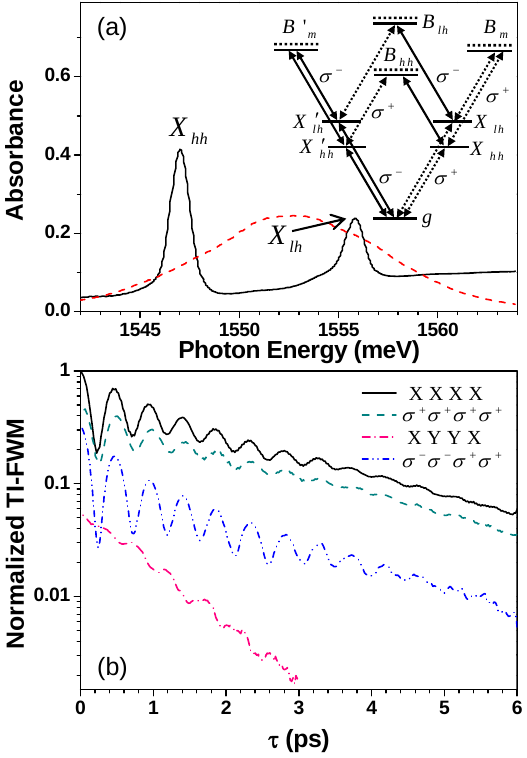}}
\caption{(Color online) (a) The linear absorption (solid black
line) and excitation laser (red dotted line) spectra. Inset shows
the level scheme for the heavy- and light-hole excitons ($X$) and
biexcitons ($B$) in GaAs quantum wells. (b) Time-integrated
four-wave mixing data for various excitation polarizations.}
\label{fig:fig2}
\end{figure}


The epitaxially grown sample consists of a four period GaAs
multiple-QW with Al$_{0.3}$Ga$_{0.7}$As barriers, where both wells
and barriers are 10 nm thick. The substrate has been removed and
all measurements are performed in transmission at approximately
7~K. Linear absorbance is shown in Fig.~\ref{fig:fig2}(a). The
peaks correspond to the heavy-hole $X_{hh}$ and light-hole
$X_{lh}$ excitons. Excitations may also include pure ($B_{hh}$ and
$B_{lh}$), as well as mixed ($B_{m}$) biexciton states, as
indicated by the level scheme in the top right corner of
Fig.~\ref{fig:fig2}(a).

Figure~\ref{fig:fig2}(b) shows the three-pulse time-integrated
TFWM signal, where $T=200$~fs and the excitation density is
$\sim8\times10^{9}$~cm$^{-2}$ per layer. The laser is tuned
between the excitons, as indicated by the red dashed line in
Fig.~\ref{fig:fig2}(a) All nonlinear data are measured in the
$\chi^{(3)}$ regime. Transients are shown for co-linear (XXXX),
co-circular ($\sigma^{+}\sigma^{+}\sigma^{+}\sigma^{+}$),
cross-linear (XYYX) and cross-circular
($\sigma^{-}\sigma^{-}\sigma^{+}\sigma^{+}$) polarized excitation.
This notation corresponds to the polarization state of the three
pump pulses ($A$, $B$ and $C$) and the emission, from left to
right. The transients are normalized to the maximum of the XXXX
polarized data (at $\tau=0$~fs). The frequency of the observed
beats is related to the splitting between $X_{hh}$ and $X_{lh}$.
For all polarizations, except XYYX, the dephasing rates are
similar. For XYYX the signal is much weaker, the dephasing is more
rapid and the beats are in anti-phase compared to the other
polarization (due to the circular selection
rules).\cite{Schmitt-Rink} For direct comparison to the rephasing
2DFT spectra the transients are shown for positive $\tau$ only.

\begin{figure}[b]
\centering{\includegraphics[width=8.5cm]{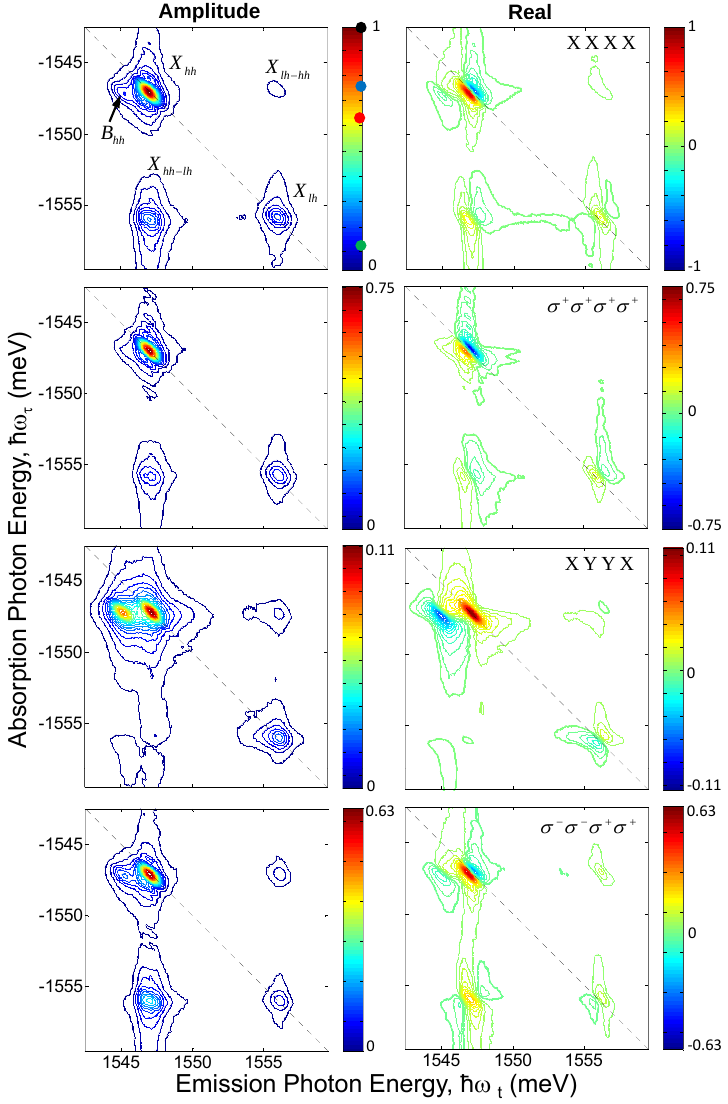}}
\caption{(color online) Amplitude and , real-part 2DFT spectra for
various polarizations are shown in the left and right panels
respectively. Each spectrum has its own colorbar and the amplitude
of the \emph{X}$_{\text{hh}}$ is marked on the top left colorbar
for the various polarizations. Vertically the rows correspond to
XXXX, $\sigma^{+}\sigma^{+}\sigma^{+}\sigma^{+}$, XYYX and
$\sigma^{-}\sigma^{-}\sigma^{+}\sigma^{+}$ polarized excitation.
All spectra are taken at a delay between the second and third
pulses of $T $= 200 fs. Note that the negative regions of the real
spectra are outlined by a thicker contour below which each
contours are dashed.} \label{fig:fig3}
\end{figure}

The amplitude and real part of the rephasing 2DFT spectra are
shown in Fig.~\ref{fig:fig3}, for the same four co- and
cross-polarizations as above. The phase-resolved spectra for
co-polarized excitations have been presented
previously.\cite{Zhang2} Here they are shown for comparison with
improved resolution with the real parts of the cross-polarized
spectra. The excitation density, laser tuning and time $T$ are the
same as those used in the TFWM data. The emission frequency is
used to determine the arithmetic sign of the frequencies, thus the
$\hbar\omega_{\tau}$ axis is negative because the first pulse is
conjugated. All spectra are normalized to the strongest peak,
namely the X$_{hh}$ peak in the amplitude spectrum for XXXX
configuration. Because the signal amplitudes vary for the
different polarization configurations, each spectrum has an
individual scale on its colorbar. To aid the quantitative
comparison dots are added to the amplitude XXXX-polarized
colorbar: the four dots represent the relative strength of the
$X_{hh}$ peak for XXXX (black dot),
$\sigma^{+}\sigma^{+}\sigma^{+}\sigma^{+}$ (blue dot),
$\sigma^{-}\sigma^{-}\sigma^{+}\sigma^{+}$ (red dot) and XYYX
(green dot) polarizations, respectively from top to bottom. The
2DFT spectra show similar contributions, as labelled in the top
left panel, including: two diagonal features associated with
$X_{hh}$ and $X_{lh}$; two off-diagonal coupling peaks
($X_{hh-lh}$ and $X_{lh-hh}$); and in some cases axial peaks
corresponding to a third-order coherence through the biexciton
state (most prominently $B_{hh}$). Diagonal elongation of the some
peaks is a sign of inhomogeneity and is unchanged by polarization.
For the $X_{hh}$ the ratio of inhomogeneous to homogeneous
broadening is approximately 3.5:1, and less for the $X_{lh}$.

The first (top) row of Fig.~\ref{fig:fig3} shows the co-linear
polarized 2DFT spectrum, where the dominant peak is the $X_{hh}$.
A weak axial $B_{hh}$ peak is observed to be about 10\% of the
$X_{hh}$, displaced towards the $\hbar\omega_{\tau}$ axis. The
real part of the spectrum shows dispersive lineshapes resulting
from EIS.\cite{Li} Additionally, the $B_{hh}$ feature is a
negative dip sitting on the shoulder of the $X_{hh}$. This
negative dip occurs because the polarization associated with the
$\left\vert X\right\rangle $ to $\left\vert B\right\rangle $
transition has the opposite sign to that for the $\left\vert
g\right\rangle $ to $\left\vert X\right\rangle $ transition.
Theory has shown that the strength of the $X_{hh}$ is enhanced and
its lineshape is dispersive due to many-body
interactions.\cite{Zhang2,Yang}

The second row of Fig.~\ref{fig:fig3} shows the co-circular
polarized 2DFT spectrum, where the strength of the $X_{hh}$ peak
and its dispersive lineshape indicate the continued dominance of
many-body interactions. This spectrum is weaker than for XXXX
excitation (approximately 75\%), because some perturbative
pathways are switched off.\cite{Yang} Pure biexcitons are
spin-forbidden, thus the $B_{hh}$ feature is absent from the
spectrum. Mixed biexcitons features are allowed and expected as
axial peaks adjacent to the $X_{hh-lh}$ and $X_{lh-hh}$ cross
peaks. The lack of mixed biexcitonic features suggest that they
have weaker oscillator strengths than the $B_{hh}$ contribution.
In addition, the off-diagonal peak strengths are weaker and more
asymmetric than for XXXX, which is consistent with the reduced
beats in the co-circular TFWM data in Fig.~\ref{fig:fig2}(b).
Modelling this spectrum has been addressed
previously,\cite{Kuznetsova} where it was found that experiments
can only be reproduced by including Coulomb correlations beyond
the Hartree-Fock approximation.\cite{Zhang2}

Row three of Fig.~\ref{fig:fig3} shows the cross-linear polarized
2DFT spectrum. The dominant features are the $X_{hh}$, $X_{lh}$,
and $B_{hh}$ peaks, where the biexcitons is a negative dip in the
real spectrum. The spectrum is significantly weaker than the
others presented (approximately 13\% that of XXXX excitation), and
the real part exhibits an absorptive lineshape for the $X_{hh}$
feature. This change in lineshape indicates a suppression of
many-body interactions, although the $X_{lh}$ lineshape remains
dispersive. Many-body interactions are suppressed for
cross-polarized excitation because there is no spatial modulation
of the net population and the excitation induced scattering
processes are supposedly spin independent. Thus, $X_{hh}$ and
$B_{hh}$ have comparable strengths and the spectrum is described
well by the perturbative pathways determined from the level
scheme.\cite{Yang} Additionally, the off-diagonal features have
almost vanished, consistent with the reduced beating in
Fig.~\ref{fig:fig2}(b).

Row four of Fig.~\ref{fig:fig3} shows the 2DFT spectra for
cross-circular polarized excitation. The $X_{hh}$ peak is about
63\% of its strength for co-linear excitation and has a dispersive
lineshape in the real part, indicating many-body interactions.
Compared to co-circular excitation both diagonal exciton features
have reduced strength, which we attribute to the transfer of
spectral weight to the $B_{hh}$ and the off-diagonal features. The
$X_{hh-lh}$ and $X_{lh-hh}$ coupling features are also more
symmetric than for co-circular excitation. This is consistent with
the cleanest beats in the TFWM data. We suspect that inter-exciton
coupling increases because scattering between excitons with the
same electron spin is allowed. This spin-dependent behavior is
elucidated by the comparison of the 2DFT spectra.

The 2DFT spectra clarify the behavior of the transients in
Fig.~\ref{fig:fig2}(b). Comparison between the two data types
show: the overall strength of the transients matches well with the
heights of the diagonal peaks of the 2DFT spectra; the relative
strength of the oscillations in the transients agree with the
strength of the off-diagonal peaks in the 2DFT spectra, indicating
that the oscillations are quantum beats and not polarization
beats. The lineshapes of the phase-resolved cross-linear 2DFT
spectra clearly indicates a suppression of many-body interactions,
which is less ambiguous than the phase-shifted beats and increased
dephasing rate of the transient. 2DFT spectra isolate the various
contributions of the TFWM, and therefore the importance of each
contribution is determined by its relative strength and 2D
lineshapes.

In summary, we have shown a set of phase-resolved 2DFT spectra for
co- and cross-polarized excitation conditions, of which the latter
were previously unattainable. The strength and lineshape of the
cross-linear polarized excitation show suppression of many-body
interactions. Since the same number of perturbative pathways
contribute to co- and cross-linear excitation it is clear that
many-body interactions dominate all other polarization
configurations. For cross-linear polarization the reduced exciton
strength is comparable to the biexciton, so they contribute
equally to the coherent response. A comparison of the two
circular-polarized spectra show a spin dependence of the
off-diagonal coupling features. The strengths and lineshapes of
these phase-resolved 2DFT spectra provide a clearer understanding
into the $\chi^{(3)}$ nonlinear response from semiconductor QWs.
Further analysis promises insight into open questions including
the fast decay for cross-polarized excitation.

$ $

This work was supported by the National Science Foundation and the Chemical
Sciences, Geosciences, and Biosciences Division Office of Basic Energy Sciences,
U.S. Department of Energy.


\begin{thebibliography}{99}

\bibitem{CundiffOpEx} S.~T.~Cundiff, Opt. Express \textbf{16}, 4639 (2008).

\bibitem{Axt1} V.~M.~Axt and T.~Kuhn, Rep. Prog. Phys. \textbf{67}, 433
(2004).

\bibitem{Cundiff1} S.~T.~Cundiff, H.~Wang and D.~G.~Steel, Phys. Rev. B
\textbf{46}, R7248 (1992).

\bibitem{Yaffe} H.~H.~Yaffe, Y.~Prior, J.~P.~Haribison and L.~T.~Florez, J.
Opt. Soc. Am. B \textbf{10}, 578 (1993).

\bibitem{Bennhardt} D.~Bennhardt, P.~Thomas, R.~Eccleston, E.~J.~Mayer and
J.~Kuhl, Phys. Rev. B \textbf{47}, 13485 (1993).

\bibitem{Saiki} T.~Saiki, M.~Kuwata-Gonokami, T.~Matsusue and H.~Sakaki,
Phys. Rev. B \textbf{49}, 7817 (1994).

\bibitem{Kim} D.-S.~Kim, J.~Shah, T.~C.~Damen, W.~Sch\"{a}fer, F.~Jahnke, S.~Schmitt-Rink
and K.~K\"{o}hler, Phys. Rev. Lett. \textbf{69}, 2725 (1992).

\bibitem{Schmitt-Rink} S.~Schmitt-Rink, D.~Bennhardt,
V.~Heuckeroth, P.~Thomas, P.~Haring, G.~Maidorn, H.~Bakker,
K.~Leo, D.-S.~Kim, J.~Shah, K.~K\"{o}hler, Phys. Rev. B
\textbf{46}, 10460 (1992).

\bibitem{Patkar} S.~Patkar, A.~E.~Paul, W.~Sha, J.~A.~Bolger and A.~L.~Smirl,
 Phys. Rev. B \textbf{51}, 10789 (1995).

\bibitem{Robart} D.~Robart, T.~Amand, X.~Marie, M.~Brousseau, J.~Barrau and
G.~Bacquet, J. Opt. Soc. Am. B \textbf{13}, 1000 (1996).

\bibitem{Bott} K.~Bott, O.~Heller, D.~Bennhardt, S.~T.~Cundiff, P.~Thomas,
E.~J.~Mayer, G.~O.~Smith, R.~Eccleston, J.~Kuhl and K.~Ploog, Phys. Rev. B
\textbf{48}, 17418 (1993).

\bibitem{Wang} H.~Wang, K.~Ferrio, D.~G.~Steel, Y.~Z.~Hu, R.~Binder, S.~W.~Koch,
 Phys. Rev. Lett. \textbf{71}, 1261 (1993).

\bibitem{Hu} Y.~Z.~Hu, R.~Binder, S.~W.~Koch, S.~T.~Cundiff, H.~Wang and D.~G.~Steel,
 Phys. Rev. B \textbf{49}, 14382 (1994).

\bibitem{Smirl} A.~L.~Smirl, M.~J.~Stevens, X.~Chen and
O.~Buccafusca, Phys. Rev. B \textbf{60}, 8267 (1999).

\bibitem{Shacklette} J.~M.~Shacklette and S.~T.~Cundiff,
Phys. Rev. B \textbf{66}, 045309 (2002).

\bibitem{Mayer} E.~J.~Mayer, G.~O.~Smith, V.~Heuckeroth, J.~Kuhl, K.~Bott,
A.~Schulze, T.~Meier, D.~Bennhardt, S.~W.~Koch, P.~Thomas, R.~Hey and K.~Ploog,
Phys. Rev. B \textbf{50}, 14730 (1994).

\bibitem{Schafer} W.~Sch\"{a}fer, D.~S.~Kim, J.~Shah, T.~C.~Damen,
 J.~E.~Cunningham, K.~W.~Goossen, L.~N.~Pfeiffer and  K.~K\"{o}hler,
Phys. Rev. B \textbf{53}, 16429 (1996).

\bibitem{Adachi} S.~Adachi, T.~Miyashita, S.~Takeyama, Y.~Takagi, A.~Tackeuchi
 and M.~Nakayama, Phys. Rev. B \textbf{55}, 1654 (1997).

\bibitem{Nickolaus} H.~Nickolaus and F.~Henneberger, Phys. Rev. B \textbf{57},
 8774 (1998).

\bibitem{Langbein} W.~Langbein and J.~M.~Hvam, Phys. Rev. B \textbf{61}, 1692 (2000).

\bibitem{Schneider} H.~Schneider and K.~Ploog, Phys. Rev. B \textbf{49}, 17050 (1994).

\bibitem{Breunig} H.~G.~Breunig, T.~Voss, I.~R\"{u}ckmann,
J.~Gutowski, V.~M.~Axt and T.~Kuhn, J. Opt. Soc. Am B \textbf{20},
1769 (2003).

\bibitem{Voss} T.~Voss, I.~R\"{u}ckmann, J.~Gutowski, V.~M.~Axt and
T.~Kuhn,Phys. Rev. B \textbf{73}, 115311 (2006).

\bibitem{Li} X.~Li, T.~Zhang, C.~N.~Borca, S.~T.~Cundiff, Phys. Rev. Lett.
\textbf{96}, 057406 (2006).

\bibitem{Langbein2} W.~Langbein and B.~Patton, J. Phys.: Condens. Matter
\textbf{19}, 295203 (2007).

\bibitem{Zhang2} T.~Zhang, I.~Kuznetsova, T.~Meier, X.~Li, R.~P.~Mirin,
P.~Thomas and S.~T.~Cundiff, Proc. Natl. Soc. USA, \textbf{104},
14227 (2007).

\bibitem{Jonas} D.~M.~Jonas, Annu. Rev. Phys. Chem. \textbf{54}, 425 (2003).

\bibitem{Bristow2} A.~D.~Bristow, D.~Karaiskaj, X.~Dai and S.~T.~Cundiff,
 Opt. Express \textbf{16}, 18017 (2008).

\bibitem{Yang} L.~Yang, I.~V.~Schweigert, S.~T.~Cundiff and S.~Mukamel,
Phys. Rev. B \textbf{75}, 125302 (2007).

\bibitem{Kuznetsova} I.~Kuznetsova, P.~Thomas, T.~Meier, T.~Zhang,
X.~Li, R.~P.~Mirin and S.~T.~Cundiff, Sol. State Comm.
\textbf{142}, 154 (2007).











\end{thebibliography}
\end{document}